\documentclass[%
aps, rmp, a4,
twocolumn, 
amsfonts, amsmath,
nofootinbib, floatfix
] {revtex4-2}
\usepackage{graphicx}
\usepackage[colorlinks=true]{hyperref}
\usepackage{framed}
\usepackage{bm}
\usepackage{ctable}
\usepackage{array}
\newcolumntype{Q}{>{$}l <{$}}
\newcolumntype{S}{>{$[}l <{]$}}
\newcolumntype{C}{>{$}c <{$}}

\usepackage{amssymb}

\def\kv{{\bf k}}
\def\rv{{\bf r}}
\def\tv{{\bf t}}
\def\cross{\!\times\!}
\def\GM{{\cal G}_M}

\begin{document}

\title{Magnetic Point Groups and Space Groups\\%
\small{Expanded and corrected draft for the Encyclopedia of Condensed Matter Physics, 2$^{\text{nd}}$ edition (Elsevier, expected 2022).}}

\author{Ron Lifshitz}
\email[Corresponding author:\ ]{ronlif@tau.ac.il}
\affiliation{Raymond and Beverly Sackler School of Physics \& Astronomy, Tel Aviv University, Tel Aviv 69978, Israel}
\affiliation{School of Mathematics, University of Leeds, Leeds LS2 9JT, United Kingdom}

\date{March 15, 2022}

\begin{abstract}

The spatial symmetry of matter---including finite objects like mole\-cules or atomic clusters, and extended objects like periodic or aperiodic crystals---is described using point groups and space groups. Magnetic point groups and space groups are the simplest extension of this description, to matter whose atomic constituents possess a property that can take one of two possible values, like the ``up'' or ``down'' orientations of a magnetic moment, or a spin. Magnetic groups---also known as antisymmetry groups, Shubnikov groups, Heesch groups, Opechowski-Guccione groups, as well as dichromatic, 2-color, or simply black-and-white groups---are here defined, and their structure and notation explained, while providing some pedagogical examples of their enumeration. The resulting magnetic selection rules, or extinctions, in neutron diffraction experiments are discussed. Further extensions to color groups and spin groups are briefly described. 

\end{abstract}


\maketitle


\section*{Keywords}

\noindent Magnetic point groups; magnetic space groups; magnetic symmetry; time reversal; antisymmetry; color groups; black-and-white groups; color symmetry; spin groups; Shubnikov groups; crystallography; crystals; periodic crystals; aperiodic crystals; quasicrystals; magnetic crystals; magnetic extinctions; magnetic Bragg peaks; Hermann-Mauguin; Opechowski-Guccione (OG); Belov-Neronova-Smirnova (BNS); Rokhsar-Wright-Mermin (RWM).

\section*{Key Points}

\begin{itemize}
  
\item The spatial symmetry of matter is described using point groups and space groups. Magnetic point groups and space groups are the simplest extension of this description to matter whose atomic constituents possess a property that can take one of two possible values.

\item Magnetic point groups are used to describe the point symmetry of  magnetically ordered objects---whether these objects are finite, like molecules or clusters, or extended, like periodic or aperiodic crystals. They are defined below, and their structure and notation explained, while providing some pedagogical examples of their enumeration. 

\item Magnetic space groups are used to describe the full magnetic symmetry of extended magnetically ordered, periodic or aperiodic, crystals. They are also defined below, and their structure and different notations explained, while providing some pedagogical examples of their enumeration. 

\item Magnetic groups are also known in the literature as antisymmetry groups, Shubnikov groups, Heesch groups, Opechowski-Guccione groups, as well as dichromatic, 2-color, or simply black-and-white groups. 

\item One of the most direct physical consequences of having magnetic symmetry in crystals, is the systematic extinction of magnetic Bragg peaks in neutron diffraction patterns. These can be calculated directly from the magnetic space group.

\item Magnetic groups are naturally generalized to cases where the property of interest is not limited to having one of only two possible values. The common extensions are to color groups with more than just two distinct colors, or to spin groups, where the orientation of the magnetic moments can vary continuously. 
  
\end{itemize}


\section{Introduction}

Magnetic groups---also known as antisymmetry groups, Shubnikov groups, Heesch groups, Opechowski-Guccione groups, as well as dichromatic, 2-color, or simply black-and-white groups---are the simplest extension to standard point group and space group theory.\footnote{The symmetry of nonmagnetic \emph{periodic} crystals is discussed in numerous crystallography textbooks. For a nice informal introduction see \citet{Senechal90}.  For a detailed overview of the symmetry of nonmagnetic \emph{quasiperiodic} crystals see \citet{mermin92}; for a more elementary introduction see \citet{Lifshitz96}.} They allow directly to describe, classify, and study the physical consequences of the symmetry of materials, characterized by having a certain added property that locally can take one of two possible values. These may include finite objects like molecules or atomic clusters, as well as extended periodic or aperiodic crystals. To do so, one introduces a single additional symmetry operation of order two that interchanges the two possible values everywhere. This operation can be applied to the material along with any of the standard point group or space group operations, and is commonly denoted by adding a prime to the original operation. Thus, any rotation $g$ followed by this external operation is denoted by $g'$.

We begin in section~\ref{sec:two} with a few typical examples of this two-valued property, some of which are illustrated in Figure~\ref{fig:chains}. In section \ref{sec:point} we discuss the notion of a magnetic point group, followed by a discussion on magnetic space groups in section \ref{sec:space}. In section \ref{sec:extinctions} we describe one of the most direct physical consequences of having magnetic symmetry in crystals, which is the systematic extinction of magnetic Bragg peaks in neutron diffraction patterns. We finish in section \ref{sec:general} by briefly describing the generalization of magnetic point groups and space groups to cases where the property of interest is not limited to having one of only two possible values.

\section{Two-valued properties}
\label{sec:two}

Let us begin by considering the structure of crystals of the cesium chloride (CsCl) type.  In these periodic crystals the atoms are located on the sites of a body-centered cubic (bcc) lattice.  Atoms of one type are located on the simple cubic lattice formed by the corners of the cubic cells, and atoms of the other type are located on the simple cubic lattice formed by the body centers of the cubic cells. The parent bcc lattice is bipartitioned in this way into two identical sublattices, related by a half body-diagonal translation.  One may describe the symmetry of the cesium chloride structure as having a simple cubic lattice of ordinary translations, with a basis containing one cesium and one chlorine atom.  Alternatively, one may describe the symmetry of the crystal as having twice as many translations per unit volume, forming a bcc lattice, half of which are primed to indicate that they are followed by an operation that exchanges the chemical identities of the two types of atoms. A similar situation occurs in crystals whose structure is of the sodium chloride (NaCl) type in which a simple cubic lattice is bipartitioned into two identical face-centered cubic (fcc) sublattices.

Another typical example is the orientational ordering of magnetic moments or spins, electric dipole moments, or any other tensorial property associated with each site in a crystal. Adopting the language of spins, if these can take only one of two possible orientations---``up'' and ``down''---as in simple Ising antiferromagnets, we have the same situation as above, with the two spin orientations replacing the two chemical species of atoms. In the case of spins, the external prime operation is a so-called antisymmetry operation that reverses the directions of the spins. Physically, one may think of using the operation of time inversion to reverse the directions of all the spins without affecting anything else in the crystal.

\begin{figure}
  \centering
  \includegraphics[width=\linewidth]{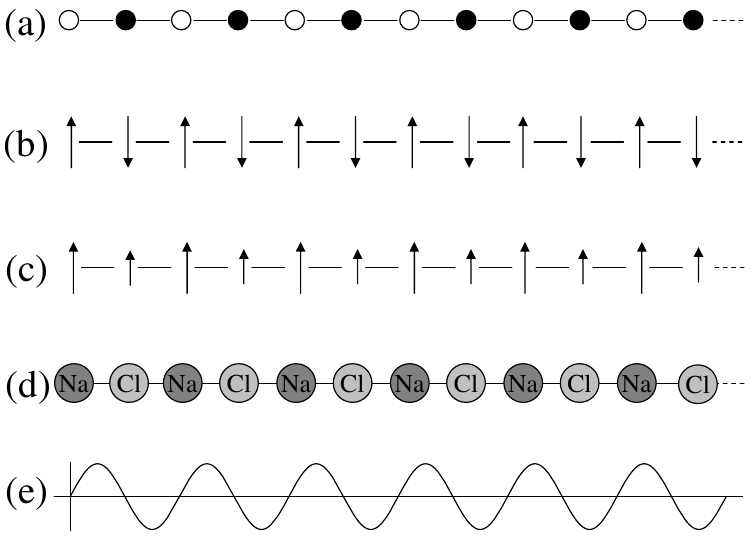}
  \caption{Several realizations of the simple one-dimensional magnetic space group $p_p\bar1$ (or $p_pm$). Note that the number of symmetry translations is doubled if one introduces an operation $e'$ that interchanges the two possible values of the property, associated with each crystal site.  Also note that spatial inversion $\bar1$ (or mirror reflection $m$) can be performed without a prime on each crystal site, and with a prime ($\bar1'$ or $m'$) between every two sites. (a) An abstract representation of the two possible values as the two colors---black and white. The operation $e'$ is the nontrivial permutation of the two colors. (b) A simple antiferromagnetic arrangement of spins. The operation $e'$ is time inversion which reverses the directions of the spins. (c) A ferromagnetic arrangement of two types of spins, where $e'$ exchanges them as in the case of two colors. (d) A 1-dimensional version of salt, where $e'$ exchanges the chemical identities of the two atomic species. (e) A function $S(x)$ whose overall sign changes by the application of the operation $e'$.}
\label{fig:chains}
\end{figure}

As a final example, consider a periodic or quasiperiodic scalar function of space, $f(\rv)$, as in Figure~\ref{fig:chains}(e), whose mean value is zero. One can often extend the point group or space group, describing the symmetry of such functions, by introducing an external prime operation that flips the sign of $f(\rv)$, by multiplying it by $-1$.

For the sake of clarity we shall adopt the picture of up-and-down spins, and the use of time inversion to exchange them, for the remaining of our discussion, but some terminology from the color description will be used when helpful. It should be emphasized, though, that most of what we say here---perhaps excluding the discussion of extinctions in section \ref{sec:extinctions}---applies equally to any other two-valued property. We describe the magnetic crystal using a scalar spin-density field $S(\rv)$ whose magnitude gives the magnitude of an atomic magnetic moment, or some coarse-grained value of the magnetic moment, at position $\rv$. The sign of $S(\rv)$ gives the direction of the spin---a positive sign for ``up'' spins and a negative sign for ``down'' spins. The function $S(\rv)$ can be a discrete set of positive or negative delta functions, defined on the crystal sites as in Figure~\ref{fig:chains}(b), or a continuous spin-density field as in Figure~\ref{fig:chains}(e).

\section{Magnetic point groups}
\label{sec:point}

A $d$-dimensional \emph{magnetic point group} $G_M$ is a subgroup of $O(d)\cross 1'$, where $O(d)$ is the orthogonal group of $d$-dimensional rotations, and $1'=\{e,e'\}$ is the time inversion group containing the identity $e$ and the time inversion operation $e'$. Note that we use the term ``rotation'' to refer to proper as well as improper rotations, such as mirrors and the spatial inversion.

Three types of subgroups exist: (1) The most trivial subgroups are simply the ordinary nonmagnetic point groups, containing rotations alone, with no use of the time-inversion operation; (2) Slightly less trivial subgroups are those in which all rotations appear both with and without time-inversion; and (3) Nontrivial subgroups are those where exactly half of the rotations are followed by time inversion and the other half are not. In the context of color symmetry, these three types of subgroups are often called \emph{colorless, gray,} and \emph{black-and-white} groups, respectively, a terminology which will become clear shortly.

More formally, if $G$ is any ordinary nonmagnetic point group, namely, any subgroup of $O(d)$, it can be used to construct at most three types of magnetic point groups $G_M$, forming its so-called \emph{magnetic superfamily}, as follows:
\begin{enumerate}
\item $G_M=G$, so that no rotation is followed by time inversion. It is an ordinary nonmagnetic point group.
\item $G_M=G\cross 1'$, so that each rotation in $G$ appears in $G_M$ once by itself and once followed by time inversion. Note that in this case $e'\in G_M$.
\item $G_M=H + g' H$, where $H$ is a subgroup  of index 2 in $G$, and $g$ is a particular element of $G$ which is not in $H$, so that $G=H + gH$. Thus, exactly half of the rotations in $G$, which belong to its subgroup $H$, appear in $G_M$ by themselves, and the other half, belonging to the coset $gH=\{gh|h\in H\}$, are followed by time inversion. Note that in this case $e'\not\in G_M$.
\end{enumerate} 

Enumeration of magnetic point groups is thus straightforward. Any ordinary point group $G$---that is, any subgroup, usually finite, of $O(d)$---is trivially a colorless magnetic point group of type 1, and along with the time inversion operation generates a gray magnetic point group of type 2, denoted as $G1'$. One then lists all distinct subgroups $H$ of index 2 in $G$, if there are any, to construct nontrivial black-and-white magnetic point groups of type 3. These are denoted either by the group-subgroup pair $G(H)$, or as we shall do here, by using the International (Hermann-Mauguin) symbol for $G$, while adding a prime to those elements in the symbol that do not belong to the subgroup $H$ and are therefore followed by time inversion.

A simple way of finding all the subgroups $H$ of index 2 in $G$ is to pick a set of generators for $G$, and consider all combinations in which each generator of \emph{even} order is either primed or unprimed. The resulting black-and-white point groups then need to be checked for equivalence as conjugate subgroups of $O(d)\times 1'$ and grouped into so-called \emph{magnetic geometric crystal classes}. As an example, we enumerate in Table~\ref{Table:MPG} the distinct classes of magnetic point groups in the orthorhombic crystal system. Tables of all magnetic point groups that are compatible with periodic crystals---and therefore restricted to rotational axes of order 1, 2, 3, 4, and 6---appear in the original work by \citet{Shubnikov64} and in the more recent book by \citet{Litvin13}, while \citet{Lifshitz97} provides the extension of these tables to all magnetic point groups in 2- and in 3-dimensions with rotational axes of arbitrary order.

\begin{table}
  \begin{ruledtabular}
    \begin{tabular}{QQQ}
      \multicolumn{3}{c}{\bf Orthorhombic Magnetic Point Groups }\\ \midrule
    \textrm{Colorless} & \textrm{Gray} & \textrm{Black \& White} \\ \midrule
    222 & 2221' & 2'2'2 \\
    mm2 & mm21' & m'm2';\ m'm'2  \\
    mmm & mmm1' & mmm';\ m'm'm;\ m'm'm' \\
  \end{tabular}
  \end{ruledtabular}
  \caption{Magnetic point groups in the orthorhombic crystal system. Each row lists the magnetic superfamily of one of the ordinary, or colorless, orthorhombic point groups $G$, listed in the first column. The second column lists the trivial gray groups $G1'$. To enumerate the black-and-white point groups---listed in the third column by their International (Hermann-Mauguin) symbol---one considers all distinct subgroups $H$ of index 2 in $G$. For example, the orthorhombic point group $G=222=\{e,2_x,2_y,2_z\}$ has three subgroups of index 2, $H_1=\{e,2_x\}$, $H_2=\{e,2_y\}$, and $H_3=\{e,2_z\}$. This gives three corresponding black-and-white point groups---$22'2'$, $2'22'$, and $2'2'2$---in which two 2-fold rotations are primed and the third is not. As these groups are all equivalent as conjugate subgroups of $O(3)\times 1'$, they are merely different settings of just a single geometric crystal class of black-and-white point groups. On the other hand, taking the two mirrors as generators of the orthorhombic point group $G=mm2=\{e,m_x,m_y,2_z\}$, one can associate time inversion with just a single one of them (in which case the 2-fold rotation is also primed), or with both. This gives three black-and-white point groups---$m'm2'$, $mm'2'$, and $m'm'2$---of which only the first two are equivalent, yielding two geometric crystal classes of black-and-white point groups. In total there are $3+3+6=12$ types of orthorhombic magnetic point groups in three dimensions.}
  \label{Table:MPG}
\end{table}

Black-and-white magnetic point groups (type 3) can be used to describe the point symmetry of finite objects, as described in Sec.~\ref{sec:finitepoint}, as well as that of infinite crystals, as described in Sec.~\ref{sec:crystalpoint}.  Trivial, or colorless, magnetic point groups (type 1), equivalent to ordinary non-magnetic point groups, are usually mentioned only for the sake of completeness. Strictly speaking, they can be said to describe the symmetry of ferromagnetic materials with no spin-orbit coupling. Gray magnetic point groups (type 2) can be used to describe the point symmetry of infinite crystals, but because they include $e'$ as a symmetry element they cannot, strictly speaking, be used to describe the symmetry of finite magnetic or two-colored objects, as explained below.

\subsection{Magnetic point groups of finite objects}
\label{sec:finitepoint}

The magnetic symmetry of a finite object, such as a molecule or a cluster containing an equal number of two types of spins, can be described by a magnetic point group. We say that a primed or unprimed rotation from $O(d)\cross 1'$ is in the magnetic point group $G_M$ of a {\it finite object\/} in $d$ dimensions, if it leaves the object {\it invariant\/}. Clearly, only black-and-white magnetic point groups (type 3 in the list above), can describe the symmetry of finite magnetically ordered structures. This is because time inversion $e'$ changes the orientations of all the spins, and therefore cannot leave the object invariant unless it is accompanied by a nontrivial rotation. For this reason, point groups of type 2 are often called ``gray'' groups, as they describe the symmetry of ``gray'' objects which remain invariant under the exchange of black and white. For a comprehensive discussion on the magnetic symmetry of finite objects, see the original work by \citet{Shubnikov64}.

\subsection{Magnetic point groups of crystals}
\label{sec:crystalpoint}

The magnetic point group of a $d$-dimensional magnetically ordered {\it periodic crystal,} containing an equal number of up and down spins, is defined as the set of primed or unprimed rotations from $O(d)\cross 1'$ that leave the crystal {\it invariant to within a translation}.  The magnetic point group of a crystal can be either of the first or of the second types listed above. It is a gray group (of type 2) if time inversion by itself leaves the crystal invariant to within a translation. In this case, any rotation in the magnetic point group can be performed either with or without time inversion. If time inversion cannot be combined with a translation to leave the crystal invariant, the magnetic point group is black and white (of type 3), in which case half of the rotations are performed without time inversion and the other half with time inversion.

\begin{figure*}
  \centering
  \includegraphics[width=0.4\textwidth]{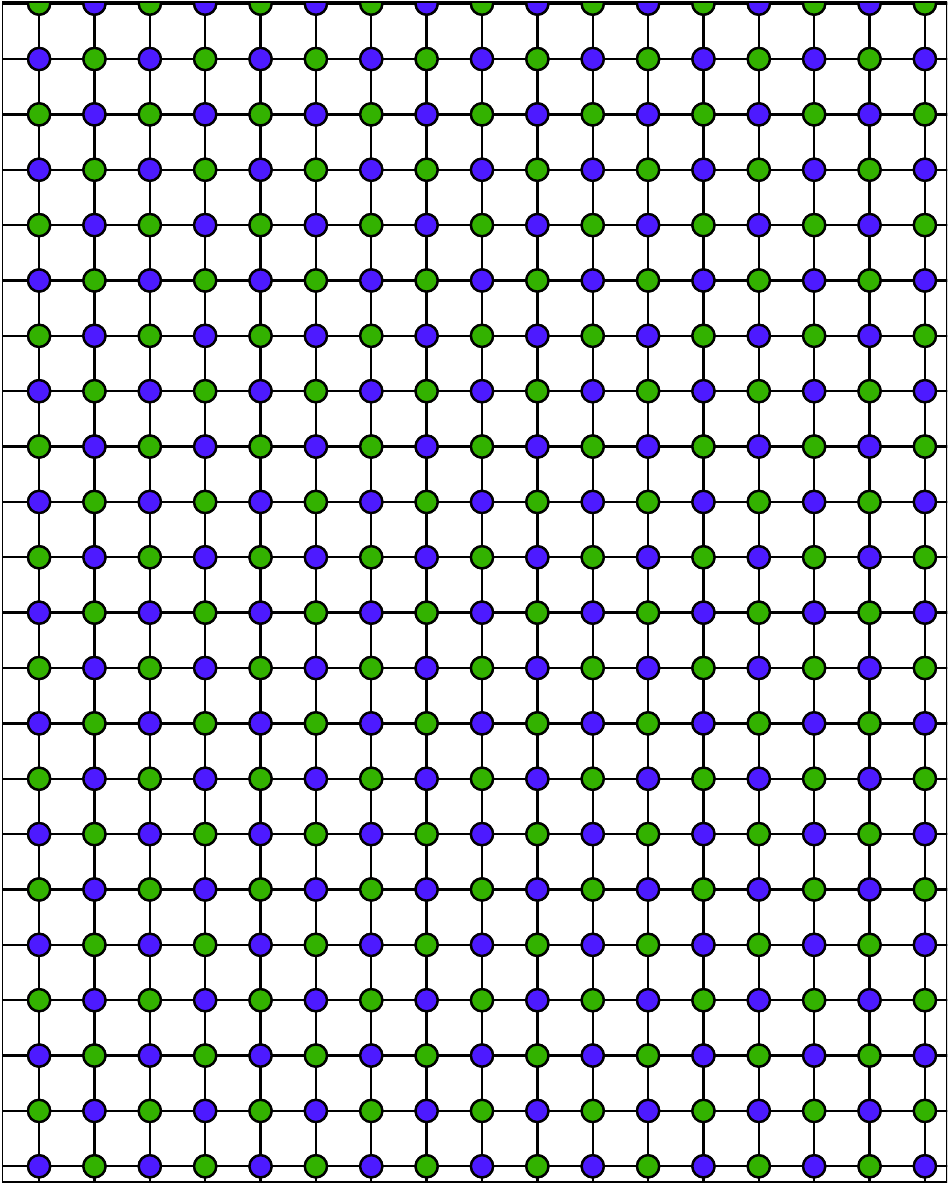}
  \includegraphics[width=0.4\textwidth]{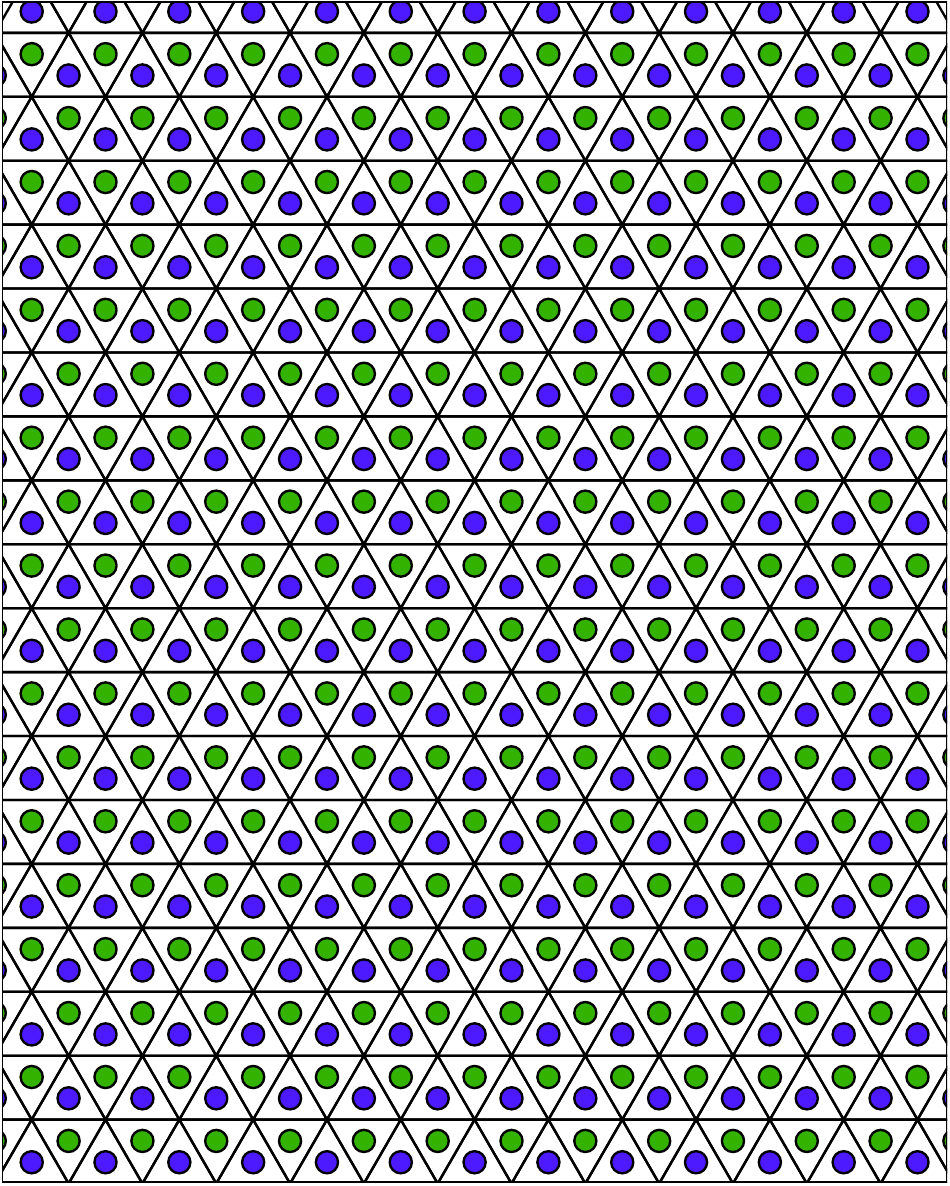}\\
(a) \hskip6.7cm (b)
\caption{\small Periodic antiferromagnets. (a) Tetragonal crystal with a gray magnetic point group $G_M=4mm1'$, and a class-equivalent magnetic space group $p_p4mm$ (also denoted $p_c4mm$). Time inversion leaves this tetragonal antiferromagnetic periodic crystal invariant to within a translation. (b) Hexagonal crystal with a black-and-white magnetic point group $G_M=6'mm'$, and a lattice-equivalent magnetic space group $p6'mm'$. Time inversion does not leave this hexagonal antiferromagnetic periodic crystal invariant to within a translation.}
\label{fig:periodic}
\end{figure*}

\begin{figure*}
\centering
\includegraphics[width=0.4\textwidth]{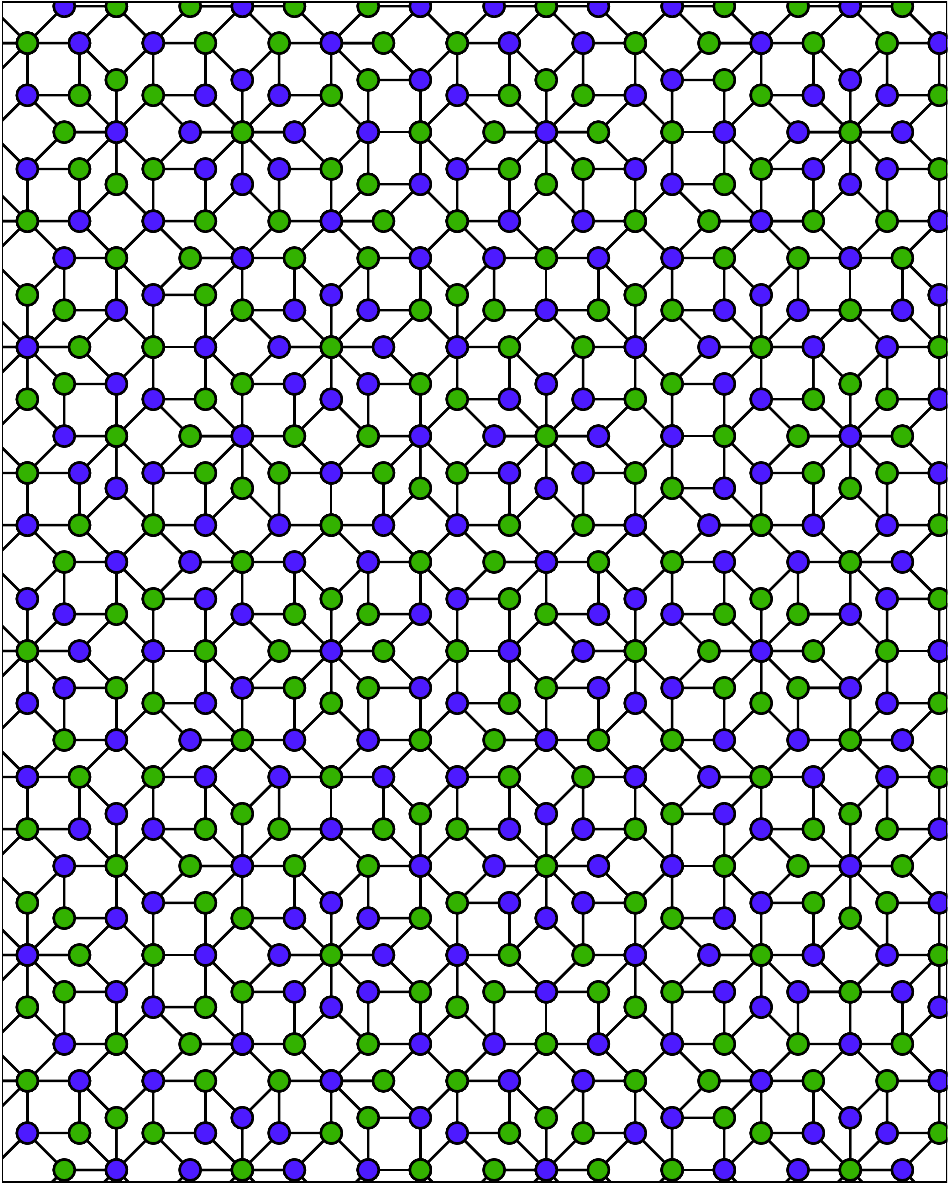}
\includegraphics[width=0.4\textwidth]{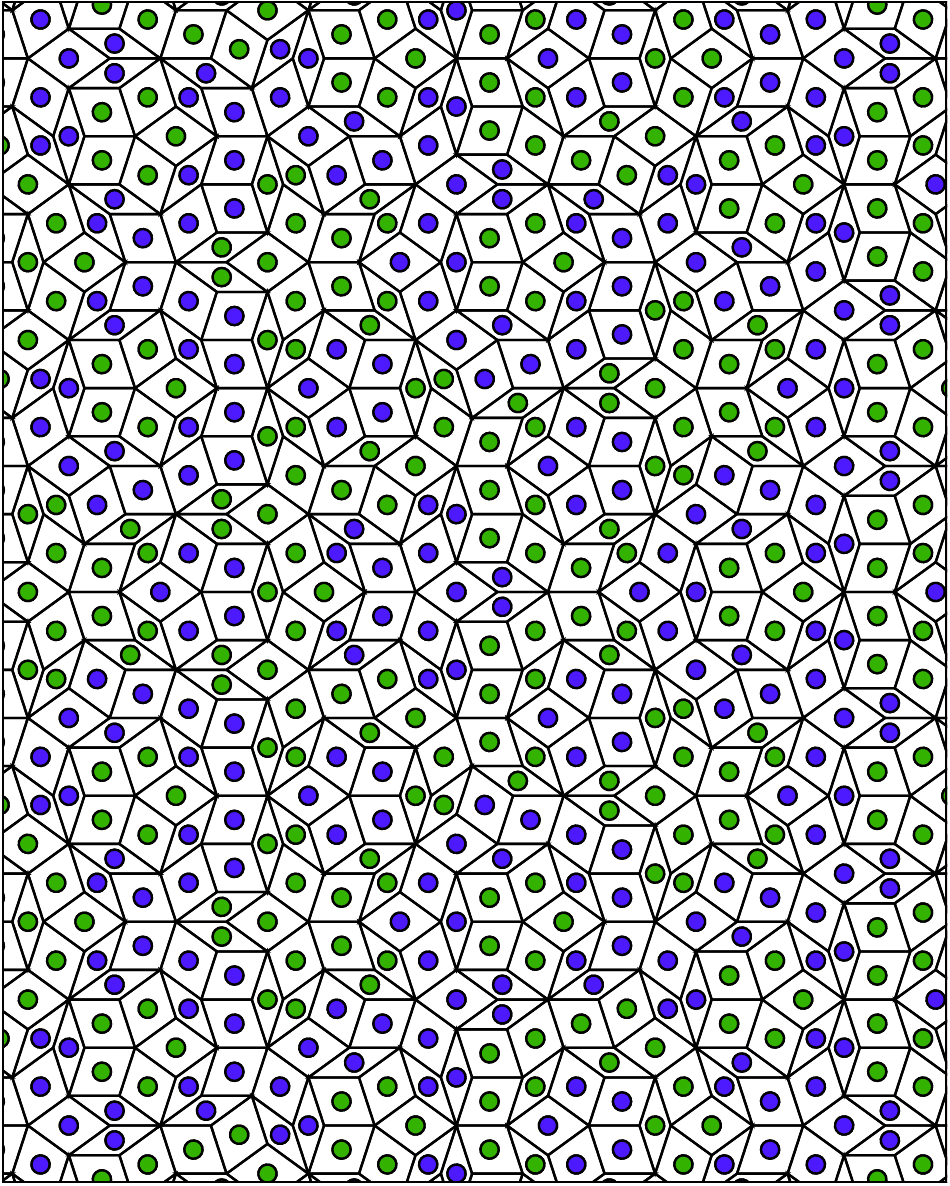}\\
(a) \hskip6.7cm (b)
\caption{Antiferromagnetic quasicrystals. (a) Octagonal quasicrystal with a gray magnetic point group $G_M=8mm1'$, and a class-equivalent magnetic space group $p_p8mm$. Time inversion leaves this octagonal antiferromagnetic quasicrystal indistinguishable. (b) Decagonal quasicrystal with a black-and-white magnetic point group $G_M=10'm'm$, and a lattice-equivalent magnetic space group $p10'm'm$. Time inversion does not leave this decagonal antiferromagnetic quasicrystal indistinguishable.}
\label{fig:QP}
\end{figure*}

Figure \ref{fig:periodic}(a) shows an example of a magnetically ordered periodic crystal, with tetragonal symmetry, given by the magnetic point group $G_M=4mm1'$, which is of type 2. Here, time inversion can be followed by a translation to leave the crystal invariant. Figure \ref{fig:periodic}(b) shows an example of a periodic crystal with hexagonal symmetry, given by the magnetic point group $G_M=6'mm'$, which is of type 3. Note that all right-side-up triangles contain a blue circle (spin up) and all up-side-down triangles contain a green circle (spin down). Time inversion, exchanging the two types of spins, cannot be combined with a translation to recover the original crystal.  Time inversion must be combined with a nontrivial operation---such as mirrors of the horizontal type or odd powers of the 6-fold rotation---that interchanges the two types of triangles to recover the original crystal. Note that mirrors of the vertical type (the first $m$ in the International symbol) leave the crystal invariant without requiring time inversion, and are therefore unprimed in the symbol.

More generally, the magnetic point group of a $d$-dimensional magnetically ordered crystal---whether periodic or not\footnote{Of particular relevance is the generalization from periodic crystals to the wider category of \emph{quasiperiodic crystals}. Quasiperiodic crystals that are explicitly aperiodic are called \emph{quasicrystals}. For proper definitions and detailed terminology, see, for example \citet{Lifshitz03, Lifshitz07, Lifshitz11}.}---is defined as the set of primed or unprimed rotations from $O(d)\cross 1'$ that leave the crystal {\it indistinguishable}. This means that the rotated and unrotated crystals contain the same spatial distributions of finite clusters of spins of arbitrary size. The two crystals are thus statistically the same, though not necessarily identical to within a translation. Only for the special case of periodic crystals does the notion of indistinguishability reduce to invariance to within a translation.

Figure \ref{fig:QP} shows two aperiodic examples, analogous to the two periodic examples of Figure \ref{fig:periodic}. Figure \ref{fig:QP}(a) shows an octagonal crystal with magnetic point group $G_M=8mm1'$. One can see that time inversion rearranges the spin clusters in the crystal but they all still appear in the crystal with the same spatial distribution. This is because any finite spin cluster and its time-reversed image appear in the crystal with the same spatial distribution. Figure \ref{fig:QP}(b) shows a decagonal crystal with magnetic point group $G_M=10'm'm$. In this case, time inversion does not leave the crystal indistinguishable. It must be combined either with odd powers of the 10-fold rotation, or with mirrors of the vertical type, to produce a crystal with the same spatial distributions of finite spin clusters as in the original one.

\section{Magnetic space groups}
\label{sec:space}

The full symmetry of a magnetically ordered crystal, described by a set of ``up'' and ``down'' spins, or a coarse-grained scalar spin-density field $S(\rv)$, is given by its magnetic space group $\GM$.  We said that the magnetic point group $G_M$ is the set of primed or unprimed rotations that leave a periodic crystal invariant to within a translation, or more generally, leave a quasiperiodic crystal indistinguishable. We still need to specify the distinct sets of translations $\tv_g$ or $\tv_{g'}$ that can accompany the rotations $g$ or $g'$ in a periodic crystal to leave it invariant, or provide the more general characterization of the nature of the indistinguishability in the case of quasicrystals, as explained below.

\subsection{Magnetic symmetry of periodic crystals}
\label{sec:periodic}

One can take any ordinary nonmagnetic space group $\cal G$ of a periodic crystal, also called a Fedorov-Sch\"{o}nflies space group, and form its \emph{magnetic superfamily}. As in the case of magnetic point groups, there are two trivial magnetic space-group types:
\begin{itemize}
\item[I.]{\it Colorless magnetic space groups}: These are ordinary nonmagnetic Fedorov-Sch\"{o}nflies space groups, where $\GM=\cal G$.
\item[II.]{\it Gray magnetic space groups}: Here $\GM={\cal G}\times 1'$ (denoted simply as  ${\cal G}1'$), implying that every operation in $\cal G$ can be applied with or without being followed by time inversion.
\end{itemize}
Nontrivial black-and-white magnetic space groups take the form
\begin{equation}\label{eq:MSG}
  \GM={\cal H} + (g'|\tv_{g'}){\cal H},
\end{equation}
where ${\cal H}$ is a subgroup of index 2 in $\cal G$, and $(g'|\tv_{g'})$ denotes a primed rotation $g'$, possibly followed by a translation $\tv_{g'}$. They are divided into two types, depending on their magnetic point groups $G_M$:
\begin{itemize}
\item[III.]{\it Lattice-equivalent magnetic space groups}: If the magnetic point group $G_M$ is black and white (of type 3), the point group $H$ of $\cal H$ is a subgroup of index 2 of the point group $G$ of $\cal G$, and the Bravais lattices of $\cal G$ and $\cal H$ are the same. Rotations in $G$ that are not in $H$ must be combined with time inversion to leave the magnetic crystal invariant to within a translation.
\item[IV.]{\it Class-equivalent magnetic space groups}: If the magnetic point group $G_M$ is gray (of type 2), with all rotations in $G$ appearing with and without time inversion, the point groups of $\cal G$ and $\cal H$ are the same, and the Bravais lattice $T$ of $\cal H$ is a sublattice of index 2 of the Bravais lattice $T_0$ of $\cal G$. Translations in $T_0$ that are not in $T$ must be combined with time inversion to leave the magnetic crystal invariant.
\end{itemize}

Enumeration of lattice-equivalent magnetic space-group types is straightforward. Given an ordinary nonmagnetic space group $\cal G$, simply consider all distinct subgroups $H$ of index 2 of the point group $G$ of $\cal G$, as explained in section~\ref{sec:point}. The only additional consideration is that there may be more than one way to orient the subgroup $H$ relative to the Bravais lattice of translations that leave the periodic crystal invariant.  Take, for example, the magnetic point group $m'm2'$, which as explained in Table~\ref{Table:MPG}, is equivalent to $mm'2'$. Consequently, on a primitive orthorhombic lattice, $Pm'm2'$ and $Pmm'2'$ are two different settings of the same magnetic space-group type. But, on an $A$-centered orthorhombic lattice, where the $x$- and $y$-axes are no longer equivalent, one needs to consider both options as distinct groups.  We say that the two distinct magnetic space-group types  $Am'm2'$ and $Amm'2'$ belong to the same magnetic \emph{geometric} crystal class ({\it i.e.}\ have the same type of magnetic point group), but to two distinct magnetic \emph{arithmetic} crystal classes.  Lattice-equivalent magnetic space groups, also called translation-equivalent when the crystal is periodic, are denoted by taking the International symbol for $\cal G$, and adding a prime to those elements in the symbol that do not belong to $\cal H$ and must be followed by time inversion.

Enumeration of class-equivalent magnetic space-group types proceeds by considering all distinct pairs of lattices $T_0$ and $T$ in a given crystal system, where $T$ is a sublattice of index 2 in $T_0$. Translations in the sublattice $T$, making up the Bravais lattice of $\cal H$, leave the periodic magnetic crystal invariant, while translations in $T_0$ that are not in $T$ must be combined with time inversion to leave it invariant. Like a color-blind person, an x-ray diffraction experiment, which is insensitive to the magnetic ordering, will exhibit the reciprocal lattice $L_0$ of $T_0$, while neutron diffraction will expose the additional Bragg peaks, contained in the reciprocal lattice $L$ of $T$, encoding the magnetic order. Correspondingly, the lattice $T_0$, and its reciprocal $L_0$, are often called the colorless or color-blind lattices, as well as the paramagnetic or disordered lattices, while the sublattice $T$, and its reciprocal $L$, are often called the magnetic or ordered lattices.\footnote{The reference to order-disorder is motivated by transitions like the one that occurs the in the $\beta$ phase of brass~\cite{Madsen16}.} Note that while $T$ is a sublattice of $T_0$ of index 2, their reciprocal lattices in Fourier space play opposite roles, whereby $L_0$ is a sublattice $L$ of index 2. 

Different settings and notations are used to describe class-equivalent magnetic space groups, depending on which of the two lattices---the magnetic or the paramagnetic---and which of the two spaces---real space or Fourier space---are considered as primary. In the \citet{Opechowski65}, or OG setting, the paramagnetic lattice of translations $T_0$ is considered primary, lending its symbol to the magnetic space group, while the magnetic lattice $T$ is denoted as a subscript. In the OG setting the class-equivalent magnetic space group is considered a member of the magnetic superfamily of the space group $\cal G$. In the Belov, Neronova, and Smirnova (\citeyear{BNS64}), or BNS setting, the magnetic lattice of translations $T$ is considered primary, lending its symbol to the magnetic space group, while the paramagnetic lattice $T_0$ is indicated as a subscript. In the BNS setting the class-equivalent magnetic space group is considered a member of the magnetic superfamily of the space group $\cal H$. Finally, if one follows the Rokhsar, Wright, and Mermin (\citeyear{RWM88}), or RWM formalism, in which the nonmagnetic space groups are viewed primarily in Fourier-space, and which generalizes most easily to aperiodic magnetic crystals~\cite[Sec.~V]{Lifshitz97}, the magnetic reciprocal lattice $L$ is considered primary, lending its symbol to the magnetic space group, while the paramagnetic reciprocal lattice $L_0$ appears as a subscript. As in the BNS setting, the class-equivalent magnetic space group is considered a member of the magnetic superfamily of the space group $\cal H$, but it is the subscript that denotes a sublattice of the main symbol as in the OG notation.

In the cubic system, for example, there are two lattice-sublattice pairs of index 2---the simple cubic lattice $P$ has a face-centered sublattice $F$ of index 2; and the body-centered cubic lattice $I$ has a simple cubic sublattice $P$ of index 2. The first pair is denoted $P_F$ in the OG notation; $F_S$ ($S$ for simple) in the BNS notation; and $I^*_P$ in the RWM notation, with the asterisk indicating that the lattice is specified in Fourier space. Recall that the reciprocal of an fcc lattice is a bcc lattice and \emph{vice versa}. The second pair is denoted $I_P$ in the OG notation; $P_I$ in the BNS notation; and $P_{F^*}$ in the RWM notation. 

In the top part of Table~\ref{Table:MSG} we enumerate all the cubic magnetic space group types with point group 432. Note that in addition to the different settings, there is a discrepancy in the different notations, which is due to the fact that when $e'\in G_M$, ordinary mirrors or rotations when unprimed may become glide planes or screw axes when primed, and \emph{vice versa.} The only consistent way to avoid such confusion is to use only unprimed elements for the symbols of class-equivalent magnetic space-group types. This is the approach adopted by both the BNS and the RWM notations, but unfortunately not by the OG notation. Also note that there is no $1'$ at the end of the symbol, as it is clear from the existence of a subscript on the lattice symbol that the magnetic point group $G_M$ contains $e'$. Comparisons between the OG and the BNS notions, and discussions regarding the advantages and disadvantages of each approach, are still ongoing, with new notations being proposed to this day \cite{Grimmer09, Grimmer10, Litvin16, Campbell22}.

\begin{table}
  \begin{ruledtabular}
    \begin{tabular}{QC | Q | Q | Q | Q | Q}
      \multicolumn{7}{c}{\bf Magnetic Space-group types}\\
      \multicolumn{7}{c}{\bf with Point Groups 432 and 532}\\
      \midrule
      \multicolumn{2}{c|}{\textrm{I.}} & \multicolumn{1}{c|}{\textrm{II.}} & \multicolumn{1}{c|}{\textrm{III.}} & \multicolumn{3}{c}{\textrm{IV.}} \\
       &\#&&& \multicolumn{1}{c|}{\textrm{BNS}} & \multicolumn{1}{c|}{\textrm{RWM}} & \multicolumn{1}{c}{\textrm{OG}} \\
      \midrule
      P432 & (\emph{207}) & P4321' & P4'32' & P_I432 & P_{F^*}432 & I_P432 \\
      P4_132 & (\emph{213}) & P4_1321' & P4'_132' & P_I4_132 & P_{F^*}4_132 & I_P4'_132'\\
      P4_232 & (\emph{208})  & P4_2321' & P4'_232' & P_I4_232 & P_{F^*}4_232 & I_P4'32'\\
      P4_332 & (\emph{212}) & P4_3321' & P4'_332' & P_I4_332 & P_{F^*}4_332 & I_P4_132\\
      F432 & (\emph{209})  & F4321' & F4'32' & F_S432 & I^*_P432 & P_F432\\
      F4_132 & (\emph{210})  & F4_1321' & F4_1'32' & F_S4_132 & I^*_P4_132 & P_F4_232\\
      I432 & (\emph{211}) & I4321' & I4'32' &  \textrm{\ \ ------} & \textrm{\ \ ------} & \textrm{\ \ ------} \\
      I4_132 & (\emph{214}) & I4_1321' & I4'_132' & \textrm{\ \ ------} & \textrm{\ \ ------} & \textrm{\ \ ------} \\
      \midrule
      P532 & & P5321' & \textrm{\ ------} & P_I532 & P_{F^*}532 & \textrm{\ \ ------} \\
      P5_132 & & P5_1321' & \textrm{\ ------} & P_I5_132 & P_{F^*}5_132 & \textrm{\ \ ------} \\
      F532 &  & F5321' & \textrm{\ ------} & F_S532 & I^*_P532 & \textrm{\ \ ------} \\
      F5_132 &  & F5_1321' & \textrm{\ ------} & F_S5_132 & I^*_P5_132 & \textrm{\ \ ------} \\
      I532 & & I5321' & \textrm{\ ------} &  \textrm{\ \ ------} & \textrm{\ \ ------} & \textrm{\ \ ------} \\
      I5_132 & & I5_1321' & \textrm{\ ------} & \textrm{\ \ ------} & \textrm{\ \ ------} & \textrm{\ \ ------} \\
    \end{tabular}
  \end{ruledtabular}
  \caption{Magnetic space-group types with point groups $432$ and $532$. Each row lists the magnetic superfamily of one nonmagnetic space group $\cal G$, listed in Column I by its International (Hermann-Mauguin) symbol and by its number from the International Tables in italics. Column II lists the trivial gray groups ${\cal G}1'$. Column III lists the lattice-equivalent space groups. Point group 432 has a single subgroup of index 2---the tetrahedral group 23---obtained by associating a prime with the 2-fold rotation axes (and consequently with the 4-fold axes as well). This gives a single lattice-equivalent magnetic space-group for each nonmagnetic space group $\cal G$. Point group $532$ has no subgroup of index 2, and therefore no lattice-equivalent magnetic space-groups.
Finally, there are two lattice-sublattice pairs in the cubic and the icosahedral systems. In both cases the simple or primitive lattice $P$ has a face-centered sublattice $F$ of index 2, and the body-centered lattice $I$ has a simple or primitive sublattice $P$ of index 2. Consequently, in the BNS setting, each nonmagnetic fcc or fci space group gives rise to a single class-equivalent magnetic space group with an fcc or fci magnetic lattice and a primitive paramagnetic lattice, and each primitive space group gives rise to a single class-equivalent magnetic space group with a primitive magnetic lattice and a bcc or bci paramagnetic lattice. There are no class-equivalent magnetic space groups with a bcc or a bci magnetic lattice. These are listed in column IV in three different notations for the cubic groups and in the BNS and RWM notations for the icosahedral groups. For the RWM notation, recall that the reciprocal of a face-centered lattice $F$ is a body-centered lattice in Fourier space, denoted by $I^*$, and \emph{vice versa}. In total there are $8+8+8+6=30$ types of magnetic space groups with point group $432$ and 6+6+4=16 with point group $532$.
}
  \label{Table:MSG}
\end{table}

Tables of periodic magnetic space-group types in two and in three dimensions are given in the early work by \citet[Table 11]{BT64}; Belov, Neronova, and Smirnova~(\citeyear[Table 10]{BNS64}); \citet{Opechowski65}; \citet{Koptsik66}; and \citet[Tables 11.2 and 17.3]{Opechowski86}; and more recently have also been made available online by \citet{Litvin13, Stokes22} and by the \citet{Bilbao}, accompanied by the review of \citet{PerezMato15}. Let us only summarize some statistics.  Starting from the 17 space-group types in 2 dimensions, also known as the 17 plane groups, one can form a total of 80 magnetic space-group types as follows: 17 colorless groups, 17 gray groups, and 46 black-and-white groups of which 20 are class-equivalent and 26 are lattice-equivalent. Starting from the 230 space-group types in 3 dimensions, one can form a total of 1651 magnetic space-group types as follows: 230 colorless groups, 230 gray groups, and 1191 black-and-white groups of which 517 are class-equivalent and 674 are lattice-equivalent.  These numbers have no particular significance other than the fact that they may seem surprisingly large.

\subsection{Magnetic symmetry of quasiperiodic crystals}

\subsubsection{Lattices and Bravais classes} 

Magnetically ordered quasiperiodic crystals, in general, possess no lattices of real space translations that leave them invariant, but they do have reciprocal lattices (or Fourier modules) $L$ that can be inferred directly from their neutron diffraction diagrams.  To be more concrete, consider a scalar spin-density field with a well defined Fourier transform
\begin{equation}
  \label{eq:fourier}
  S(\rv)=\sum_{\kv\in L} S(\kv) e^{i\kv\cdot\rv},
\end{equation}
in which the set $L$ contains at most a countable infinity of wave vectors. In fact, the {\it magnetic reciprocal lattice\/} $L$ of a magnetic crystal can be defined as the set of all integral linear combinations of wave vectors $\kv$, determined from its neutron diffraction diagram. Indeed, we expect to see a magnetic Bragg peak at every $\kv$ in the closure of those that are experimentally observed, unless it is forbidden by symmetry, as explained in section~\ref{sec:extinctions} below, or unless it is too weak to be detected experimentally.

The {\it rank\/} $D$ of $L$ is the smallest number of vectors needed to generate $L$ by integral linear combinations. If $D$ is finite the crystal is quasiperiodic. If, in particular, $D$ is equal to the number $d$ of spatial dimensions, and $L$ extends in all $d$ directions, the crystal is periodic. In this special case the magnetic lattice $L$ is reciprocal to a lattice $T$ of translations in real space that leave the magnetic crystal invariant, without applying time inversion.  If $D>d$ the crystal is aperiodic, and the field $S(\rv)$ describes a magnetic quasicrystal. The reciprocal lattices $L$ of magnetic crystals are classified into Bravais classes, just like those of non-magnetic crystals.

\subsubsection{Phase functions and space-group types} 

We proceed by following the Rokhsar, Wright, and Mermin (\citeyear{RWM88}, RWM) formalism and their notion of \emph{indistinguishability}, used earlier to define the magnetic point group. The precise mathematical statement of this notion in the case of magnetic symmetry \cite{Lifshitz97, Lifshitz98} is the requirement that any symmetry operation of the magnetic crystal leave invariant all spatially averaged $n^{th}$-order autocorrelation functions of its spin-density field,
\begin{equation} \label{corr}
  C^{(n)}(\rv_1,\ldots,\rv_n) =
   \lim_{V\to\infty}{1\over V}\int_V \textrm{d}\rv 
   S(\rv_1-\rv)\cdots S(\rv_n-\rv).
\end{equation} 
One can prove \cite[in the Appendix]{Lifshitz97} that two quasi\-periodic spin-density fields $S(\rv)$ and $\hat S(\rv)$ are indistinguishable in this way, if their Fourier coefficients, defined in Eq.~(\ref{eq:fourier}), are related by
\begin{equation}  \label{eq:gauge}
  \hat S(\kv) = e^{2\pi i\chi(\kv)} S(\kv),
\end{equation}
where $\chi$ is a real-valued linear function (modulo integers) on $L$, called a {\it gauge function.} By this we mean that for every pair of wave vectors $\kv_1$ and $\kv_2$ in the magnetic lattice $L$,
\begin{equation}
  \label{eq:linear}
  \chi(\kv_1+\kv_2) \equiv \chi(\kv_1) + \chi(\kv_2),
\end{equation}
where ``$\equiv$'' means equal to within adding an integer. Thus, for each element $g$ or $g'$ in the magnetic point group $G_M$ of the crystal, which by definition leaves the crystal indistinguishable, there exists a corresponding gauge function $\Phi_g(\kv)$ or $\Phi_{g'}(\kv)$, called a {\it phase function,} satisfying
\begin{equation}
  \label{eq:pointcond}
  S(g\kv) = \begin{cases}
    e^{2\pi i\Phi_g(\kv)} S(\kv), & g\in G_M,\\
    e^{2\pi i\Phi_{g'}(\kv)}[-S(\kv)], &g'\in G_M.
    \end{cases}
\end{equation}

Since for any $g,h\in G$, $S([gh]\kv)=S(g[h\kv])$, the corresponding phase functions for elements in $G_M$, whether primed or not, must satisfy a {\it group compatibility condition,}
\begin{equation}
  \label{eq:GCC}
  \Phi_{g^*h^\dagger}(\kv) \equiv \Phi_{g^*}(h\kv) + \Phi_{h^\dagger}(\kv),
\end{equation}
where the asterisk and the dagger denote optional primes. A {\it magnetic space group,} describing the symmetry of a magnetic crystal, whether periodic or aperiodic, is thus given by a magnetic (reciprocal) lattice $L$, a magnetic point group $G_M$, and a set of phase functions $\Phi_g(\kv)$ and $\Phi_{g'}(\kv)$, satisfying the group compatibility condition (\ref{eq:GCC}).

In particular, if the point group is gray, the phase function $\Phi_{e'}(\kv)$ encodes all the information regarding the sublattice $L_0$, earlier referred to as the paramagnetic or color-blind lattice. To see this, note that the relation $(e')^2=e$ implies---through the group compatibility condition (\ref{eq:GCC}) and the fact that $\Phi_e(\kv)\equiv0$---that $\Phi_{e'}(\kv)\equiv0$ or $1/2$. It then follows from the linearity of the phase function \eqref{eq:linear} that on exactly half of the wave vectors in $L$, $\Phi_{e'}(\kv)\equiv 1/2$, and on the remaining half, $\Phi_{e'}(\kv)\equiv0$. The latter half form the sublattice $L_0$, which is of index 2 in $L$.

Magnetic space groups are classified in this formalism into {\it magnetic space-group types\/} by organizing sets of phase functions satisfying the group compatibility condition (\ref{eq:GCC}) into equivalence classes.  Two such sets $\Phi$ and $\hat\Phi$ are equivalent if: (I) they describe indistinguishable spin-density fields related as in (\ref{eq:gauge}) by a gauge function $\chi$; or (II) they correspond to alternative descriptions of the same crystal that differ by their choices of absolute length scales and spatial orientations.  In case (I) $\Phi$ and $\hat\Phi$ are related by a {\it gauge transformation},
\begin{equation}
  \label{eq:gauge-tr}
  {\hat\Phi}_{g^*}(\kv) \equiv \Phi_{g^*}(\kv) + \chi(g\kv-\kv),
\end{equation}
where, again, the asterisk denotes an optional prime.

In the special case that the crystal is periodic ($D=d$) it is possible to replace each gauge function by a corresponding $d$-dimensional translation $\tv$, satisfying $2\pi\chi(\kv)=\kv\cdot\tv$, thereby reducing the requirement of indistinguishability to that of invariance to within a translation.  Only then does the point group condition (\ref{eq:pointcond}) become
\begin{equation}
  \label{eq:periodicpointcond}
  S(g\rv) =
  \begin{cases}
      S(\rv-\tv_g), & g\in G_M,\\
    -S(\rv-\tv_{g'}), & g'\in G_M,
  \end{cases}
\end{equation}
the gauge transformations (\ref{eq:gauge-tr}) turn into the so-called Frobenius congruences~\cite[page 69]{Senechal90}, which are related to simple shifts of the origin, and the whole description reduces to that given in section~\ref{sec:periodic} for periodic crystals. 

\begin{figure*}
\centering
\includegraphics[width=0.48\textwidth]{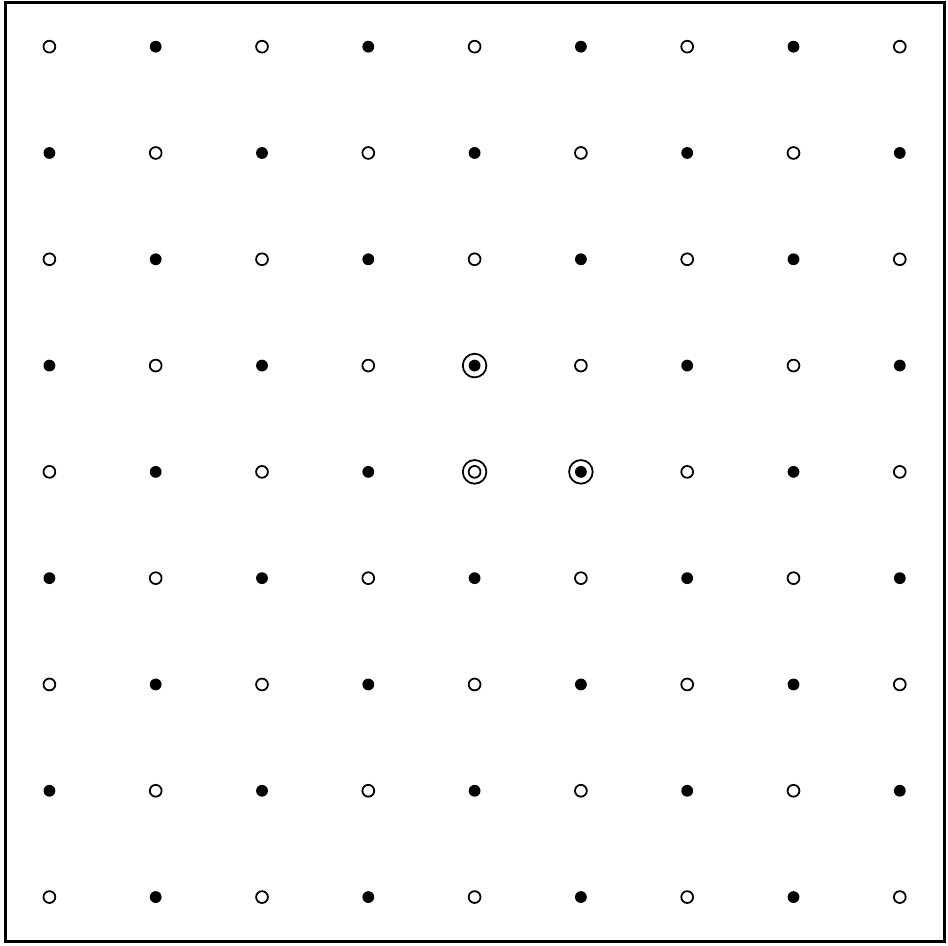}
\includegraphics[width=0.48\textwidth]{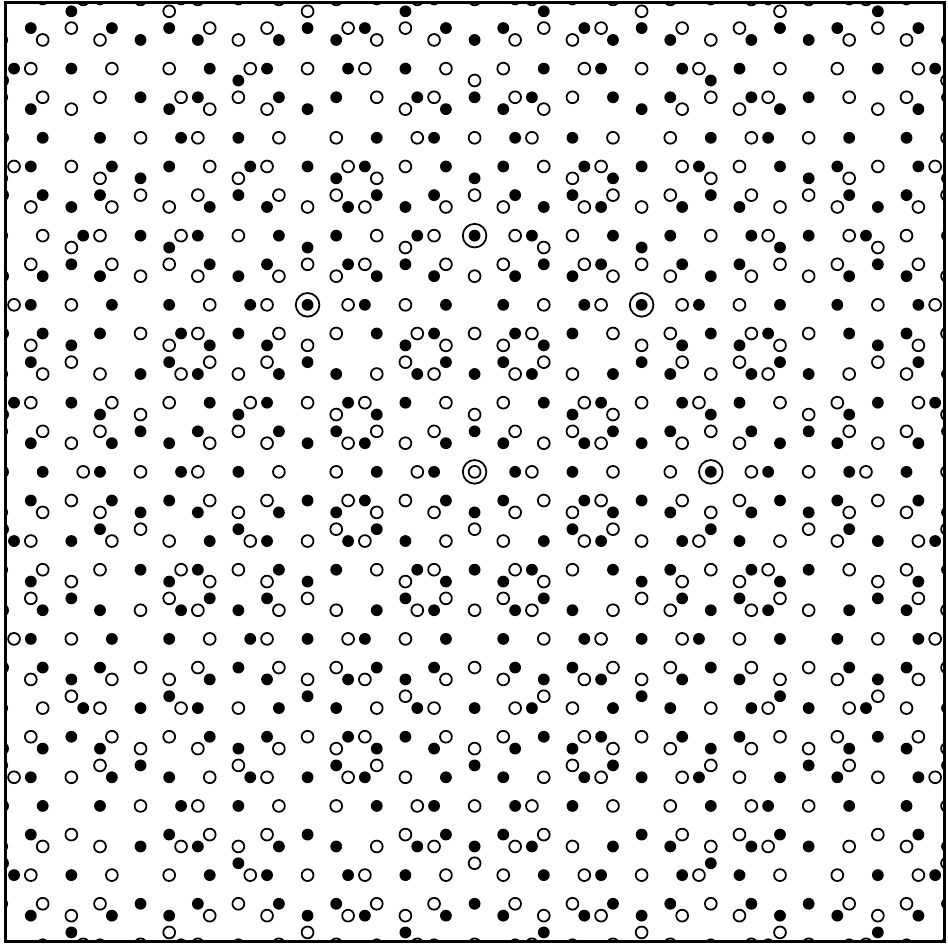}\\
(a) \hskip8.2cm (b)
\caption{\small Extinctions of magnetic Bragg peaks in crystals with class-equivalent magnetic space groups (type IV). Both figures show the positions of expected magnetic Bragg peaks indexed by integers ranging from -4 to 4. Filled circles indicate observed peaks and open circles indicate positions of extinct peaks. The origin and the $D$ vectors used to generate the patterns are indicated by an additional large circle. (a) corresponds to the tetragonal crystal in Figure~\ref{fig:periodic}(a) with magnetic space group $p_p4mm$ and (b) corresponds to the octagonal crystal in Figure~\ref{fig:QP}(a) with magnetic space group $p_p8mm$.}
\label{fig:extinctions}
\end{figure*}

As an example, we enumerate in Table~\ref{Table:MSG} the icosahedral magnetic space group types with point group 532, contrasting them with those of the cubic point group 432. Tables of magnetic space-group types on standard\footnote{See \citet[footnote 8]{Lifshitz97} for the meaning of ``standard'' in this context.} axial quasicrystals in two and three dimensions with arbitrary rotational symmetry, as well as  magnetic space-group types in the icosahedral crystal system in three dimensions, are given by \citet[Sec.~V]{Lifshitz97}.

\section{Extinctions in neutron diffraction of antiferromagnetic crystals}
\label{sec:extinctions}

We said earlier that every wave vector $\kv$ in the lattice $L$ of a magnetic crystal is a candidate for a diffraction peak unless symmetry forbids it. We are now in a position to understand exactly how this happens. Given a wave vector $\kv\in L$ we examine all magnetic point-group operations $g$ or $g'$ for which $g\kv=\kv$.  For such elements, the point-group condition (\ref{eq:pointcond}) reduces to
\begin{equation}
  \label{eq:extinctioncond}
  S(\kv) = \begin{cases}
  e^{2\pi i\Phi_g(\kv)} S(\kv), & g\in G_M ,\\
  -e^{2\pi i\Phi_{g'}(\kv)} S(\kv), & g'\in G_M,
\end{cases}
\end{equation} 
requiring $S(\kv)$ to vanish unless $\Phi_g(\kv)\equiv 0$ if $g\in G_M$, and $\Phi_{g'}(\kv)\equiv 1/2$ if $g'\in G_M$. It should be noted that the phase values in Eq.~(\ref{eq:extinctioncond}), determining the extinction of $S(\kv)$, are independent of the choice of gauge~(\ref{eq:gauge-tr}), and are therefore uniquely determined by the magnetic space-group type of the crystal, regardless of the choice of gauge.

Particularly striking are the extinctions when the magnetic space group $\cal G$ is class-equivalent (type IV), implying that the magnetic point group $G_M$ is gray (type 2), containing time inversion $e'$. Equation~\eqref{eq:extinctioncond} requires $S(\kv)$ to vanish on the whole paramagnetic sublattice $L_0$, where $\Phi_{e'}(\kv)\equiv0$. Thus, in magnetic crystals with class-equivalent space groups (type IV), at least half of the diffraction peaks are missing. Figure~\ref{fig:extinctions} shows the positions of the expected magnetic Bragg peaks corresponding to the tetragonal and octagonal magnetic structures shown in Figures~\ref{fig:periodic}(a) and \ref{fig:QP}(a), illustrating this phenomenon. Further details on magnetic extinctions including tabulated results are given by \citet{Lifshitz98, Lifshitz00, Lifshitz04, EvenDar04}.

\section{Outlook: Generalizations of magnetic groups}
\label{sec:general}

There are two natural generalizations of magnetic groups, going beyond what is discussed here. One is to color groups~\cite{Schwarzenberger84, Senechal88, Lifshitz97} with more than two distinct colors, and the other is to spin groups~\cite{Litvin73,*Litvin74,*Litvin77, Lifshitz98, Lifshitz04, *EvenDar04, Lifshitz00} where the spins are viewed as classical axial vectors free to rotate continuously in any direction.

An $n$-color point group $G_C$ is a subgroup of $O(d)\cross S_n$, where $S_n$ is the permutation group of $n$ colors. Elements of the color point group are pairs $(g,\gamma)$ where $g$ is a $d$-dimensional (proper or improper) rotation and $\gamma$ is a permutation of the $n$ colors. As before, for $(g,\gamma)$ to be in the color point group of a finite object it must leave it invariant, and for $(g,\gamma)$ to be in the color point group of a crystal it must leave it indistinguishable, which in the special case of a periodic crystal reduces to invariance to within a translation. To each element $(g,\gamma)\in G_C$ corresponds a phase function $\Phi_g^\gamma(\kv)$, satisfying a generalized version of the group compatibility condition~(\ref{eq:GCC}). The color point group contains an important subgroup of elements of the form $(e,\gamma)$ containing all the color permutations that leave the crystal indistinguishable without requiring any rotation $g$.

A spin point group $G_S$ is a subgroup of $O(d)\cross SO(d_s)\cross 1'$, where $SO(d_s)$ is the group of $d_s$-dimensional proper rotations operating on the spins, and $1'$ is the time inversion group as before. Note that the dimension of the spins need not be equal to the dimension of space ({\it e.g.} one may consider a planar arrangement of 3-dimensional spins). Also note that because the spins are axial vectors there is no loss of generality by restricting their rotations to being proper. Elements of the spin point group are pairs $(g,\gamma)$ where $g$ is a $d$-dimensional (proper or improper) rotation and $\gamma$ is a spin-space rotation possibly followed by time inversion. Here as well, elements of the form $(e,\gamma)$ play a central role in the theory, especially in determining the symmetry constraints imposed by the corresponding phase functions $\Phi_e^\gamma(\kv)$ on the patterns of magnetic Bragg peaks, observed in elastic neutron-diffraction experiments.

\section*{Acknowledgments}

I would like to thank David Mermin for teaching me how to think about crystals and their symmetry, and Shahar Even-Dar Mandel for joining me in my investigation of the magnetic symmetry in quasicrystals. I also thank Alastair Rucklidge for his kind hospitality at the School of Mathematics at the University of Leeds, where this manuscript was prepared, as well as the Cheney Foundation for their support of my visit. My research in these matters is funded by the Israel Science Foundation, most recently through grant No.~1667/16.

\bibliography{BWgroups}

\end{document}